\documentclass[preprint2,twoside]{hwo}

\usepackage{graphicx}
\usepackage{multirow}

\usepackage{rotating}
\usepackage{array}
\usepackage{booktabs}

\newcolumntype{C}[1]{>{\centering\arraybackslash}m{#1}}

\renewcommand{\arraystretch}{1.5}

\usepackage{xcolor}
\definecolor{rev}{HTML}{E82A06}
\definecolor{todo}{HTML}{FC0090}
\definecolor{grey}{HTML}{727272}

\newcommand{\grey}[1]{\textcolor{grey}{#1}}

\newcommand{\izwi}{I~Zw~18}

\input{hwo.h}

\begin{document}

\title{\textbf{\LARGE Echoes of the First Stars: Massive Star Evolution in Extremely Metal-Poor Environments with the Habitable Worlds Observatory}}
% List all primary authors here. Contributing authors may be placed here
% or in a section below, at the discretion of the primary author.
% Please include the email address for the corresponding author.
\author {\textbf{\large Peter Senchyna,$^{1}$ Calum Hawcroft,$^2$ Miriam Garcia,$^3$ Aida Wofford,$^4$  Janice C.\ Lee, $^{2}$ Chris Evans$^5$}}
\affil{$^1$\small\it The Observatories of the Carnegie Institution for Science, 813 Santa Barbara Street, Pasadena CA 91101, USA; \email{psenchyna@carnegiescience.edu}}
\affil{$^2$\small\it Space Telescope Science Institute, 3700 San Martin Drive, Baltimore, MD, 21218, USA}
\affil{$^3$\small\it Centro de Astrobiología (CAB), CSIC-INTA, Carretera de Ajalvir km 4, E-28850 Torrejón de Ardoz, Madrid, Spain}
\affil{$^4$\small\it Instituto de Astronomía, Universidad Nacional Autónoma de México, Unidad Académica en Ensenada, Km 103 Carr. Tĳuana-Ensenada, Ensenada 22860, México}
\affil{$^5$\small\it European Space Agency (ESA), ESA Oﬃce, Space Telescope Science Institute, 3700 San Martin Drive, Baltimore, MD 21218, USA}

% Please add the names of additiona contributing authors in the format "Joseph Jensen (Utah Valley University), " separated by commas.
\author{\small{\bf Contributing Authors:} 
Ryan J.\ Rickards Vaught (STScI), Macarena G.\ del Valle-Espinosa (STScI), Grace Telford (Princeton, Carnegie Observatories), Fabrice Martins (LUPM, Univ.\ Montpellier, CNRS), G\"{o}ran \"{O}stlin (Stockholm University), M\'{e}d\'{e}ric Boquien (Universidad de Antofagasta), Paul Scowen (NASA/GSFC), John Ziemer (JPL), Jane Rigby (NASA/GSFC), Sophia Flury (University of Edinburgh), Jorick Vink (Armagh Observatory and Planetarium), Swara Ravindranath (NASA GSFC, CRESST II CUA)
% List additional contributing authors here, in order of their contribution. 
% Contributing Author (affiliation), Contributing Author (affiliation), Contributing Author (affiliation)
}

% Please add the names of endorsers in the format "Joseph Jensen (Utah Valley University), " separated by commas.
\author{\footnotesize{\bf Endorsed by:}
% List endorsers in alphatical order here. Endorser 1 (affiliation), Endorser 2 (affilation), Endorser 3 (affiliation)
Karla Z. Arellano-Cordova (University of Edinburgh), Narsireddy Anugu (Georgia State University), Lingesh B (Dwaraka Doss Goverdhan Doss Vaishnav College), David Barckhoff (University of Pittsburgh), Danielle Berg (The University of Texas at Austin), Matheus Bernini-Peron (Universtät Heidelderg), Stéphane Blondin (Aix Marseille Univ/CNRS/LAM), Julia Bodensteiner (University of Amsterdam), Jean-Claude Bouret (Laboratoire d'Astrophysique de Marseille), Dominic Bowman (Newcastle University, UK), Alex Cameron (University of Oxford), Luca Casagrande (Australian National University), Hsiao-Wen Chen (The University of Chicago), Annalisa Citro (University of Minnesota), Melanie Crowson (American Public University), Paul Crowther (University of Sheffield), Alexandre David-Uraz (Central Michigan University), Annalisa De Cia (European Southern Observatory [ESO]), Jeremy Drake (Lockheed Martin Solar and Astrophysics Laboratory), Christi Erba (STScI), Sophia Flury (University of Edinburgh), Luca Fossati (Space Research Institute, Austrian Academy of Sciences), Emma Friedman (NASA GSFC), Alex Fullerton (Space Telescope Science Institute), Hongwei Ge (Yunnan Observatories Chinese Academy of Sciences), Maude Gull (UC Berkeley), Krtička Jiří (Masaryk University), Logan Jones (STScI), Josiek Joris (ZAH/ARI, Universität Heidelberg), Gloria Koenigsberger (UNAM Instituto de Ciencias Físicas), Iva Krtičková (Masaryk University), Ariane Lançon (Observatoire astronomique de Strasbourg - France), Eunjeong Lee (EisKosmos, Inc.), Roel Lefever (ZAH/ARI, Universität Heidelberg), Ilya Mandel (Monash University), Stephan McCandliss (Johns Hopkins University), Drew Miles (Caltech), Pranav Nalamwar (University of Notre Dame), Donatas Narbutis (Institute of Theoretical Physics and Astronomy, Faculty of Physics, Vilnius University, Vilnius, Lithuania), Faraz Nasir Saleem (Egypt Space Agency (EgSA)), Gijs Nelemans (Radboud University), Salafia Om Sharan (INAF - Osservatorio Astronomico di Brera), Lidia Oskinova (Potsdam University, Germany), Mardav Panwar (Meerut College, Meerut), Gioia Rau (NSF), Swara Ravindranath (NASA GSFC, CRESST II CUA), Danilo Rocha (Observatório Naceional - Brazil), Julia Roman-Duval (STScI), Andreas Sander (ZAH/ARI, Universität Heidelberg), Masrrat Siddiqui  (Bansal college of pharmacy), Linda Smith (Space Telescope Science Institute), Frank Soboczenski (University of York \& King's College London), Heloise Stevance (University of Oxford), Robert Szabo (HUN-REN CSFK Konkoly Observatory), Dorottya Szécsi (Nicolaus Copernicus University, Poland), David Thilker (Johns Hopkins University), Jorick Vink (Armagh Observatory and Planetarium), Asif ud-Doula (Penn State Scranton), Andrew Wetzel (University of California, Davis)
}

% This section is for ADS Processing.  There must be one line per author. Leave them commented out for the present. They will be included later.
% \paperauthor{Peter~Senchyna}{psenchyna@carnegiescience.edu}{0000-0002-9132-6561}{The Observatories of the Carnegie Institution for Science}{}{Pasadena}{CA}{91101}{USA}
% \paperauthor{Calum~Hawcroft}{chawcroft@stsci.edu}{0000-0003-0145-8964}{Space Telescope Science Institute}{}{Baltimore}{MD}{21218}{USA}
% \paperauthor{Miriam~Garcia}{mgg@cab.inta-csic.es}{0000-0002-7881-0754}{CSIC-INTA}{Centro de Astrobiología (CAB)}{Madrid}{}{}{Spain}
% \paperauthor{Aida~Wofford}{awofford@astro.unam.mx}{0000-0001-8289-3428}{Universidad Nacional Autónoma de México, Unidad Académica en Ensenada}{Instituto de Astronomía}{Ensenada}{Tijuana-Ensenada}{22860}{México}
% \paperauthor{Janice~C.~Lee}{jlee@stsci.edu}{0000-0002-2278-9407}{Space Telescope Science Institute}{}{Baltimore}{MD}{21218}{USA}
% \paperauthor{Chris~Evans}{chevans@stsci.edu}{0009-0005-7288-6407}{European Space Agency}{STScI ESA Office}{Baltimore}{MD}{21218}{USA}

% Please provide entries for the Author index; leave them commented out for now.
%\aindex{Senchyna, P.}
%\aindex{Hawcroft, C.}
%\aindex{Garcia, M.}
%\aindex{Wofford, A.}
%\aindex{Lee, J.C.}
%\aindex{Evans, C.}

\begin{abstract}
    A remarkable span of frontier astrophysics, from gravitational-wave archaeology to the origin of the elements to interpreting snapshots of the earliest galaxies, depends sensitively on our understanding of massive star formation and evolution in near-pristine, relatively enriched gas.
    From the surprisingly massive black holes detected by LIGO/Virgo to highly ionized nebulae with peculiar enrichment patterns observed in galaxies at Cosmic Dawn, evidence is mounting that our understanding of massive-star populations at very low metallicity remains critically incomplete.
    The fundamental limitation is the hand nature has dealt us: only a few star-forming galaxies within $\lesssim~1$ Mpc can currently be resolved into individual stars, and none reach the extreme metallicities and star-formation intensities that characterized the early Universe.
    With an ultraviolet integral-field spectrograph aboard the Habitable Worlds Observatory (HWO), this barrier will finally be broken.
    HWO will bring rare, actively star-forming, extremely metal-poor dwarf galaxies at $\sim$10--20~Mpc such as I Zw 18 within reach of resolved UV-optical spectroscopy, providing our first direct, statistical view of individual massive stars and the feedback they drive at $>$30 M$_{\odot}$ and $<$10 \% Z$_{\odot}$.
    This science is deeply synergistic with many next-generation facilities, yet requires the unique combination of spatial resolution and UV/optical sensitivity that only HWO can provide.
    The massive star science enabled by HWO within the Local Volume represents a transformational advance in our ability to probe the earliest stellar populations -- those that seeded the Milky Way and other galaxies with the first heavy elements, and paved the way for life in the transparent, reionized Universe we inhabit today.
\end{abstract}

\vspace{2cm}

\section{Introduction}
\begin{figure*}[ht!]
    \centering
    \includegraphics[width=1.0\textwidth]{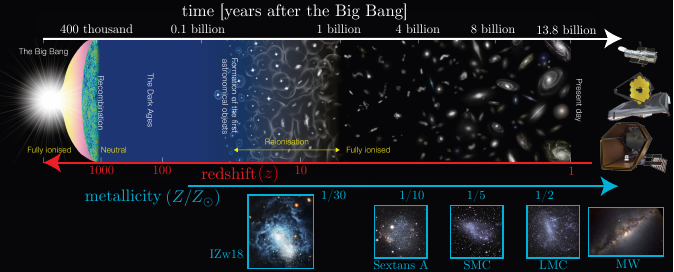}
    \caption{
        Schematic representing the chemical evolution of the Universe, and the path towards observing massive stars approaching early Universe metallicities. 
        At present, our only detailed constraints for massive stars across a broad mass range are in the Small and Large Magellanic Clouds, at $\gtrsim 20$\% solar metallicity ($Z_\odot$) - corresponding roughly to bulk metallicities characteristic of only the most recent few-billion years \citep[e.g.][]{madauCosmicStarFormationHistory2014}.
        Pioneering programs with HST have provided a glimpse of lower-metallicity stars in a handful of dwarf irregular galaxies out to the edge of the Local Group, but these only host a small number of massive stars uniformly below 50~$M_\odot$ and extending only to $\sim 10\%$ solar. 
        A facility like HWO is required to capture the first picture of a representative sample of more massive stars at metallicities actually representative of the early (the first Gyr after the Big Bang, at redshift $z>6$) Universe. (Figure adapted from NAOJ by M.~Garcia and P.~Senchyna)
        \label{fig:fig1}
    }
\end{figure*}

Understanding the origins of the first galaxies and stars is one of the great frontiers of modern astrophysics.
With the James Webb Space Telescope (JWST), we are now peering deeper into the early Universe than ever before, uncovering a surprising population of luminous galaxies in place only a few hundred million years after the Big Bang \citep[e.g.][see schematic in Figure~\ref{fig:fig1}]{adamoFirstBillionYears2024,starkObservationsFirstGalaxies2025}.
Yet, while JWST has revolutionized our ability to identify and begin to characterize these ancient systems in formation, our understanding of these observations is constrained by a major blind spot: the physics of the massive, metal-poor stars that dominate their light and which reshaped the early Universe, laying the foundations for subsequent generations of stars and galaxies including our own.

Massive stars --- those more than $\sim$8 and up to (at least) 100–300~$M_\odot$ --- are key agents of evolution in the Universe.
They flood their surroundings with ionizing radiation, launch gas flows, drive turbulence and structure in the interstellar medium through powerful winds and supernovae, and synthesize essential $\alpha$-elements such as oxygen, silicon, and calcium.
As progenitors of core-collapse supernovae, gamma-ray bursts, black holes, and compact binaries that merge to produce gravitational waves, they underlie a tremendous range of astrophysical processes --- from the origin of elements and the evolution of galaxies, to fundamental tests of gravity and spacetime.

Critically, the integrated light of the highest-redshift galaxies observed by JWST is dominated by massive stars --- both from their photospheres and the prominent surrounding nebular gas emission powered by their ionizing radiation \citep[e.g.][]{kewleyUnderstandingGalaxyEvolution2019,eldridgeNewInsightsEvolution2022}.
Interpreting these observations therefore depends directly on robust models for the atmospheres and evolution of massive stars, particularly at low metallicity.

The challenge is that the metal-poor massive stars of the early Universe differ dramatically from those we have been able to study individually in the Milky Way, Magellanic Clouds, and Local Group.
With much lower metallicities, and therefore far less efficient cooling in star-forming clouds, such environments may systematically produce more massive stars, leading to a top-heavy stellar initial mass function \citep[e.g.][]{omukaiThermalFragmentationProperties2005}.
Once formed, these stars are predicted to shed far less mass through line-driven winds, which would profoundly alter their subsequent evolution \citep[e.g.][]{kudritzkiLinedrivenWindsIonizing2002,smithMassLossIts2014,vinkTheoryDiagnosticsHot2022,martinsSpectroscopicEvolutionVery2022}.
Under these conditions, interactions with close binary companions may play an important role, producing hot stripped or rejuvenated stars whose radiation can even dominate the integrated spectra of galaxies under certain model assumptions \citep[e.g.][]{gotbergImpactStarsStripped2019,eldridgeNewInsightsEvolution2022,marchantEvolutionMassiveBinary2024}.

\begin{figure}[ht!]
    \centering
    \includegraphics[width=0.5\textwidth]{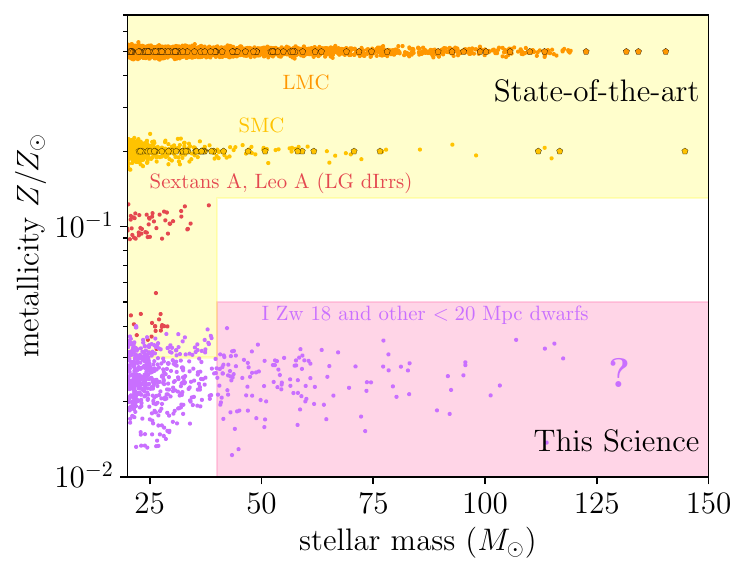}
    \caption{
        A schematic overview of the metal-poor massive star populations available to resolved spectroscopy at present and with HWO.
        The present state-of-the-art consists only of the Large and Small Magellanic Clouds \citep[L/SMC, including stars from ULLYSES:][]{bestenlehnerXShootingULLYSESMassive2025}, and a handful of metal-poor dwarf Irregular galaxies with relatively low star formation rates and consequently lacking in higher mass stars; see Table~\ref{tab:galaxies} and Section~\ref{sec:physical_parameters}.
        This science case is motivated by the lack of resolved star constraints at very low metallicities ($<10\%$ solar) and high stellar masses ($\gtrsim 30$--50~$M_\odot$; pink box), an area of parameter space into which the stars that dominate the integrated light and feedback of the earliest galaxies and those responsible for surprisingly heavy black holes found through LIGO, etc are most likely to fall.
    \label{fig:metmass}
    }
\end{figure}

Our empirical constraints on these processes remain severely limited, and models diverge substantially in their predictions \citep[e.g.][]{eldridgeSpectralPopulationSynthesis2009, gotbergImpactStarsStripped2019, klenckiPartialenvelopeStrippingNucleartimescale2022, lecroqNebularEmissionYoung2023,hovis-afflerbachMassDistributionStars2025}.
At metallicities representative of the early Universe \citep[$< 10\%$ solar, particularly in iron;][Figure~\ref{fig:fig1}]{arellano-cordovaFirstLookAbundance2022,rhoadsFindingPeasEarly2023,vanzellaExtremeIonizingProperties2024,stantonNIRVANDELSSurveyStellar2024,willottSearchFirstStars2025}, we have \emph{no} resolved spectroscopy of individual massive stars with $\gg 30$~$M_\odot$ like those that dominate the ionizing radiation and feedback in high-redshift galaxies (despite pioneering observational efforts out to edge of the Local Group, Figure~\ref{fig:metmass}).
Only recently have the first statistical samples of confidently binary-modified massive stars been identified at any metallicity, 
and surprises have arisen even in these samples reaching down only to $\sim$20\% solar in the Small Magellanic Cloud \citep[e.g.][]{schootemeijerCluesScarcityStrippedenvelope2018,droutDiscoveryMissingIntermediatemass2023,gotbergStellarPropertiesObserved2023,shenarMassiveHeliumStar2023,ramachandranXShootingULLYSESMassive2024}.

The outstanding uncertainties in the initial mass function, wind physics, and binary interaction outcomes at low metallicities have cascading effects that extend well beyond our understanding of massive stars themselves.
They directly influence the rates and properties of supernovae, the formation of compact object binaries detectable in gravitational waves, and the overall chemical and radiative feedback that regulates star and galaxy evolution.
Key properties --- such as the strength of line-driven winds and the feedback they inject into their surroundings --- remain poorly constrained in the regime most relevant to the first galaxies \citep[e.g.][]{smithMassLossIts2014,vinkTheoryDiagnosticsHot2022,martinsSpectroscopicEvolutionVery2022}.
The foundational uncertainties in metal-poor massive star models complicate our understanding of every higher-order process, from ionizing photon escape to the star formation cycle, and underlie the puzzling properties JWST is now revealing in early galaxies: unusually hard ionizing spectra, extreme emission-line ratios, and abundance anomalies \citep[e.g.][]{starkSpectroscopicDetectionIV2015,mainaliEvidenceHardIonizing2017,schmidtGrismLensAmplifiedSurvey2017,bunkerJADESNIRSpecSpectroscopy2023,toppingMetalpoorStarFormation2024,toppingDeepRestUVJWST2025,castellanoJWSTNIRSpecSpectroscopy2024}.
These are observations that current models struggle to reproduce, but that echo signatures long seen nearby in metal-poor star-forming galaxies \citep[e.g.][]{garnettHeIIEmission1991,thuanHighIonizationEmissionMetaldeficient2005,shiraziStronglyStarForming2012,senchynaExtremelyMetalpoorGalaxies2019,bergIntenseIVHe2019,olivierCharacterizingExtremeEmission2022}, in pollution by the products of hot nuclear burning in massive stars \citep[e.g.][]{cameronNitrogenEnhancements4402023,senchynaGNz11ContextPossible2024,charbonnelNenhancementGNz11First2023,marques-chavesExtremeNemittersHigh2024,toppingDeepRestUVJWST2025,vinkVeryMassiveStars2023}, and in globular clusters and other ancient stellar systems \citep[e.g.][]{bastianMultipleStellarPopulations2018,renziniTransientOvercoolingEarly2023,belokurovNitrogenEnrichmentClustered2023}.

To resolve this impasse, we need direct, spatially resolved spectroscopy of massive stars in environments that are the closest analogues to those of the early Universe.
A rare subset of nearby dwarf galaxies host young stellar populations at $<10\%$ solar metallicity \citep[e.g.][]{kunthMostMetalpoorGalaxies2000}.
However, even the nearest extremely metal-poor systems with sizeable massive star populations remain beyond the reach of current facilities in both spatial resolution and depth.
Pathfinding work out to the edge of the Local Group ($\sim1$~Mpc) has revealed hints of surprising physics among the relatively limited massive star populations within the nearest dwarf irregular galaxies below SMC metallicity \citep[e.g.][]{urbanejaAraucariaProjectLocal2008,castroARAUCARIAProjectGridbased2012,hosekQuantitativeSpectroscopyBlue2014,gullPanchromaticStudyMassive2022,lorenzoNewReferenceCatalogue2022,telfordIonizingSpectraExtremely2023,telfordObservationsExtremelyMetalPoor2024,gullLowMetallicityMassive2024, urbanejaMetallicityDistanceLeo2023,fureyWindPropertiesOtype2025,mintzSpectroscopicSurveyMetalPoor2025}, underscoring the critical importance of resolved star observations in this metallicity regime.
However, pushing well beyond the Local Group to higher-mass and lower-metallicity stars like those that dominate ionization and feedback in early galaxies requires a next-generation ultraviolet observatory.

The Habitable Worlds Observatory (HWO), a ``Super-Hubble" with unprecedented sensitivity and ultraviolet capabilities, will deliver the first resolved spectroscopy of massive stars in the critical extremely low-metallicity regime.
No other planned facility, including the ELTs, will be capable of making these measurements.
These observations will supply the missing empirical foundation for models of stellar winds, feedback, and evolution at low metallicity --- essential not only for interpreting JWST’s discoveries of the first galaxies, but also for advancing our understanding of stellar death, chemical enrichment, feedback-driven galaxy evolution, and the formation pathways of compact objects across cosmic time.

\section{Science Goals}
Here we provide a summary of the broad, overarching goals that this science case will address, including key outstanding questions identified by the Astro2020 Decadal Surveys.

Overarching science questions:

\begin{itemize}
            \item How did the earliest generations of metal-poor massive stars form and evolve?
            \item How did the stellar populations that shaped the first galaxies differ from those living in our own Milky Way now?
            \item In what ways do very low metallicities impact on the evolution of massive stars? What are the relative roles of wind-driven mass loss, binary mass transfer, rotation, and other effects in shaping massive stars in the early Universe?
            \item How does the mechanical feedback and ionizing radiation powered by massive stars change at very low metallicities?
            \item What are the yields as a function of time released by winds, supernovae, etc.\ by very metal-poor stellar populations, and how are these processed into the ISM and CGM of early galaxies?
            \item How does metallicity and its detailed effects on massive star physics manifest collectively in the integrated properties of young stellar populations?
\end{itemize}

Related Astro2020 questions:

        \begin{itemize}
            \item What are the most extreme stars and stellar populations?
            \item How does multiplicity affect the way a star lives and dies?
            \item Is the Stellar Initial Mass Function Universal?
        \end{itemize}

Relation to other fundamental science cases:
        \begin{itemize}
            \item Understanding \ion{H}{1} and \ion{He}{2} ionizing fluxes, the role in cosmic reionization, nucleosynthesis, stellar evolution, effects on the ISM (ionization, enrichment, triggered star formation, dust formation/destruction), effects on planet formation, and fate (potential supernovae and remnants) of extremely low-metallicity massive stars.
            \item Nature/origin of the first stellar-mass black hole seeds; how did massive LIGO binary BHs form?  How were earliest stellar mass black holes related to IM/SMBH seeds?
        \end{itemize}

\section{Science Objectives}

\begin{table*}
\centering
\footnotesize
\caption{Science Objectives, Physical Parameters, and Observables for Metal-Poor Massive Star Science \label{tab:objectives}}
\begin{tabular}{l|ccc}
\hline
Broad Objective & Specific Objective & Physical Parameters & Observables \\

\hline
\multirow{14}{*}{\parbox{3cm}{Constrain feedback powered by metal-poor ($\ll 10\%$~$Z_\odot$) massive stars ($>30$~$M_\odot$)}} & \multirow{6}{*}{\parbox{3cm}{Constrain the winds of individual massive stars in I~Zw~18}} & \multirow{6}{*}{\parbox{4cm}{\centering terminal wind velocity ($v_\infty$), mass loss rate ($\dot{M}$)}} & \multirow{6}{*}{\parbox{5.5cm}{\centering FUV lines including key resonant lines (\ion{O}{6}~$\lambda\lambda 1032,1038$, \ion{S}{4}~$\lambda\lambda 1062,1072$, \ion{P}{5}~$\lambda\lambda 1118,1128$, \ion{N}{5}~$\lambda\lambda 1238,1242$, \ion{Si}{4}~$\lambda\lambda 1393,1402$, \ion{C}{4}~$\lambda\lambda 1548,1550$, \ion{He}{2}~$\lambda 1640$, \ion{N}{4}~$\lambda 1718$, \ion{O}{5}~$\lambda 1371$, ...)}} \\ 
\\
\\
\\
\\
\\\cline{2-4}
 & \multirow{8}{*}{\parbox{3cm}{Map ionized gas across I~Zw~18}} & \multirow{8}{*}{\parbox{4cm}{\centering \ion{H}{1}, \ion{He}{1}, \ion{He}{2} ionizing photon production rates ($Q_{0,1,2}$), ionizing photon escape fraction, ionization parameter ($\log U$), electron temperature and density ($T_e$, $n_e$), dust extinction (stellar and nebular)}} & \multirow{8}{*}{\parbox{5.5cm}{\centering UV(--optical) nebular emission lines (\ion{C}{4}~$\lambda\lambda 1548,1550$, \ion{He}{2}~$\lambda 1640$, \ion{O}{3}]~$\lambda\lambda 1661,1666$, \ion{C}{3}]~$\lambda\lambda 1907,1909$, [\ion{Ne}{5}]~$\lambda 3426$, [\ion{O}{2}]~$\lambda\lambda 3727,3729$, [\ion{O}{3}]~$\lambda 4363$, H$\beta$~$\lambda 4861$, [\ion{O}{3}]~$\lambda\lambda 4959,5007$, ...), nebular continuum including free-bound (inc.\ the Balmer jump at 3645~\AA{}) and 2-photon emission (1216--2000+~\AA{})}} \\
\\
\\
\\
\\
\\
\\
\\
\hline
\multirow{16}{*}{\parbox{3cm}{Constrain the fundamental properties \& evolution of metal-poor ($\ll 10\%$~$Z_\odot$) massive stars ($>30$~$M_\odot$)}} & \multirow{10}{*}{\parbox{3cm}{\centering Place individual stars in I~Zw~18 in evolutionary context}} & \multirow{10}{*}{\parbox{4cm}{stellar luminosity ($L$), effective temperature ($T_{\rm{eff}}$), surface gravity ($\log g$), projected rotational velocity ($v \sin i$), HeCNO surface abundances ($Y_{He}$, $\epsilon_{C}$, $\epsilon_{N}$, $\epsilon_{O}$)}} & \multirow{10}{*}{\parbox{5.5cm}{\centering $L$: (spectro-photometry) $\lambda$1000-2000+. $T_{\rm{eff}}$(UV): Fe~III (1360-1380) -- IV (1530-1570,1600-1630,1710-1730) -- V(1440-1470)-VI(1260-1290) + C~III~1176--IV~1169 balance; $T_{\rm{eff}}$(Opt): He~I-II balance (4471,4542). log$g$: optical H lines (4340, 4860, 4100, etc). $v$sin$i$: UV/Opt metal/He lines as for $T_{\rm{eff}}$. $Y_{He}$: UV/Opt He lines.  $\epsilon$: UV/Opt metal lines. }} \\
\\
\\
\\
\\
\\
\\
\\
\\ \cline{2-4}
 & \multirow{6}{*}{\parbox{3cm}{Detect the products and signatures of binary evolution and rotation on massive stars in I~Zw~18, e.g.\ stripped stars}} & \multirow{6}{*}{\parbox{4cm}{\centering binary fraction ($\mathcal{F}_{\rm{bin}}$), wind+evolutionary parameters for evolutionary products as allowed above}} & \multirow{6}{*}{\parbox{5.5cm}{\centering FUV excess and lines (\ion{N}{5}~$\lambda \lambda 1238,1242$, \ion{He}{2}~$\lambda 1640$), optical He lines for more detailed characterization (as above).}} \\
\\
\\
\\
\\
\\ 
\bottomrule

\end{tabular}
\tablecomments{With sufficient S/N spectra, $T_{\rm{eff}}$(UV) can be constrained within 1500~K, while $T_{\rm{eff}}$(Opt) can be within 1000~K. Additional optical He lines include He--~$\lambda$~3188, 3203, 4026, 4387, 4713, 4920 - 4200, 5412~\AA{} \citep{bouretMassiveStarsSmall2021,evansNearUVReconnaissanceMetalpoor2023}. Abundance diagnostics of CNO include: C--~$\lambda$~4267, 4645-4650, 5696, 5800-5815, 6575-6585~\AA{}; N--~$\lambda$~3995, 4058, 4379, 4510-4515, 4605-4620, 4630-4640, 5200~\AA{}; O--~$\lambda$~4075, 4132, 4661, 5592~\AA{}.}
\end{table*}

In this section we outline the specific science objectives to be pursued in addressing the above goals.
Table~\ref{tab:objectives} provides a summary of these objectives.

\textit{Broad Objective:} First detailed constraints on massive star evolution and feedback under early-Universe conditions.

\textit{Specific Objective:} To benchmark models for stellar populations in the early Universe, we will spatially-resolve and measure fundamental parameters of individual massive ($>30$~$M_\odot$) stars and their ionized surroundings in extremely metal-poor environments ($<10\%$ solar metallicity, or $Z_\odot$).

This science case builds upon many decades of detailed work on the resolved massive star populations of the Milky Way and Magellanic Clouds and other Local Group systems \citep[e.g.][]{masseyMASSIVESTARSLOCAL2003, bresolinVLTSpectroscopyBlue2007,evansVLTFLAMESTarantulaSurvey2011,simon-diazIACOBProjectRotational2014,bergerQuantitativeSpectroscopySupergiants2018,shenarBinarityLOwMetallicity2024}, and examines what would be required to extend this work to a statistical sample of massive stars at significantly lower metallicities.
We split the key objectives of this case into two broad categories of physics to be constrained for extremely metal-poor massive stars: feedback (winds, mechanical energy injection by supernovae, and ionizing radiation), and evolution (fundamental properties, impact of mass loss and mass transfer, etc.).

\begin{figure}[t]
    %\centering
    \includegraphics[width=0.5\textwidth]{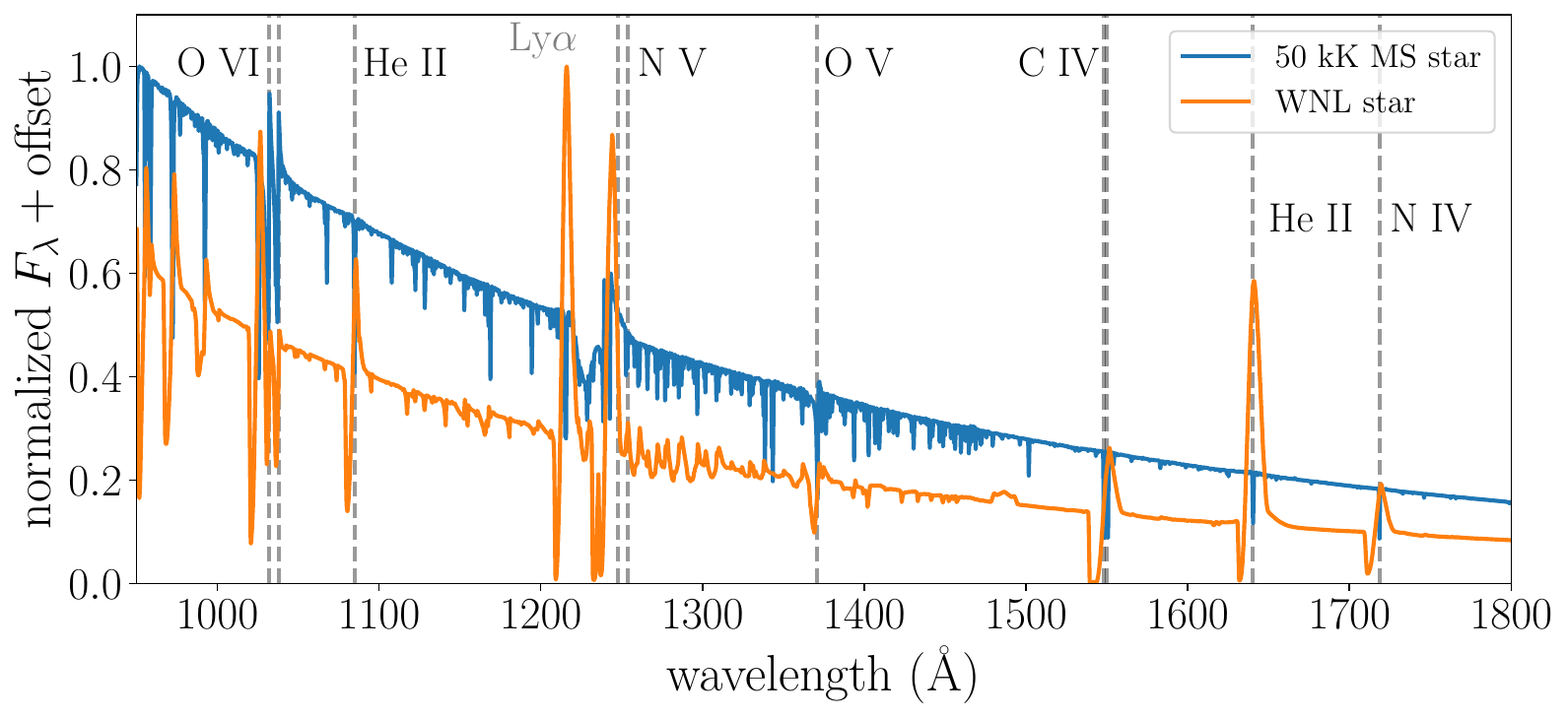}
    \caption{
        The far-ultraviolet provides access to a suite of lines formed in the atmospheres of massive stars which are required to understand the winds they drive, especially at very low metallicities where these winds are almost completely hidden at other wavelengths.
        Here we plot theoretical atmosphere models convolved to $\mathcal{R}\sim 4000$ for a hot main sequence star \citep{martinsSpectroscopicEvolutionMassive2021} and a WNL star driving a dense Wolf Rayet wind \citep{todtPotsdamWolfRayetModel2015}, both at $\lesssim 7\%~Z_\odot$.
        \izwi{} is likely to harbor stars spanning the range of wind properties represented here.
        Many of these same lines (particularly \ion{C}{4} and \ion{He}{2} in addition to others in this range) are known to be excited also in nebular gas in \izwi{} and other extremely metal-poor galaxies.
        HWO IFU spectroscopy will simultaneously provide our first glimpse of individual massive stellar atmospheres at such low-metallicity, and constrain their ionizing impact via this extended high-ionization nebular emission.
    \label{fig:uvspec}
    }
\end{figure}

\textit{(i)} Constrain the energetic feedback of massive stars at $<10\%$ solar metallicity:
\begin{itemize}
    \item Place the first empirical constraints on the photospheres and winds of a population of \emph{individual} massive stars $>30$~$M_\odot$ at extremely low metallicity: requires \emph{spatially-resolved} FUV spectroscopy.
        \begin{itemize}
            \item UV spectroscopy ($\sim$1000--1800~\AA{}) probes the brightest part of the SEDs of hot stars and provides access to key signatures sensitive to their winds and mass loss rates and photospheric abundances (e.g., Figure~\ref{fig:uvspec}).
            \item Broader wavelength coverage extending from the FUV--blue optical (including 3000--5000~\AA{}) and atmosphere modeling constrains more robustly a swath of fundamental stellar properties for comparison to evolutionary models; including a modeled estimate of ionizing radiation output.
            \item Spatial resolving power is central to this science. In galaxies or regions where massive stars are blended with neighbors, individual diagnostics become inaccessible and degeneracies result in uncertainties in the derived parameters. Even so, spectroscopy of partially resolved young clusters containing high-mass stars at the lowest metallicities will still provide new insights and advance the field.
        \end{itemize}
    \item Constrain the spatially-resolved ionizing and mechanical feedback powered by extremely metal-poor massive star populations
        \begin{itemize}
            \item Spatially-resolved FUV--optical spectroscopy of nebular gas surrounding young stellar populations is sensitive to the ionizing spectrum of the nearby stellar population (including both hot stars and e.g.\ X-ray binaries) and the impact of mechanical feedback.
            \item Mapping this gas out in emission alongside the resolved stellar populations themselves allows this \emph{feedback} (energy, momentum, chemical elements, and ionizing radiation injected into the ISM) to be directly tied to the massive stars responsible.
            \item The UV is also critical here: UV lines are uniquely sensitive to the most highly ionized and highest density gas --- tracing the most energetic and uncertain components of feedback (including cores of most highly-ionized nebulae, fast radiative supernovae shocks in dense ISM, etc.)
                \begin{itemize}
                    \item Rest-FUV lines are also some of the only tracers we have for ionized gas in highest-redshift galaxies with JWST and the ELTs: we already know these lines are notably strong in high-z galaxies and in low-metallicity systems nearby, but challenging to interpret without an understanding of where in metal-poor galaxies they are excited and by what sources.
                \end{itemize}
        \end{itemize}
\end{itemize}

\textit{(ii)} Probe the evolution and fates of massive stars at $<10\%$ solar metallicity:
\begin{itemize}
    \item Populate an empirical spectroscopic HR diagram for extremely metal-poor massive stars for the first time, for direct comparison to state-of-the-art evolutionary predictions including a range of uncertain physics (from varying mass loss prescriptions to fast rotation and binary evolution)
    \item Search directly for evidence of impacts from binary evolution (in addition to quantitative comparison to model evolutionary tracks: rotation and photospheric composition constraints, emission lines in the Balmer series formed in decretion disks, RV measurements across multiple visits or longer exposures, spectrophotometric modeling of full FUV-optical SED)
    \item Constrain the top of the IMF in extremely low-metallicity environments for the first time: place limits on the presence of very massive stars ($>100$~$M_\odot$) both among resolved stars and in the densest semi-resolved cluster cores through UV-optical spectroscopic signatures \citep[e.g.][]{martinsInferringPresenceVery2023}.
\end{itemize}

\section{Physical Parameters}
\label{sec:physical_parameters}

We subdivide the key physical parameters whose measurement is necessary to accomplish the above science into two categories: those associated with the massive stars themselves, and those associated with the ionized gas surrounding them (see Table~\ref{tab:objectives}).

\paragraph{Stellar parameters and measurements:}

\begin{itemize}
    \item Key physical parameters of (semi-)resolved stars:
    \begin{itemize}
        \item Effective temperatures, luminosities, surface gravities, projected rotation rates ($v \sin i$), terminal wind velocities ($v_\infty$), wind acceleration profiles ($\beta$), mass loss rates ($\dot{M}$), He abundances, CNO abundances, Fe abundances, microturbulence ($\xi_t$), macroturbulence ($v_\mathrm{macro}$)
    \end{itemize}
    \item For unresolved clumps/clusters: broader constraints with more degeneracies on cluster masses and ages, and some constraints on the most luminous / highest temperature components brightest in the FUV continuum and/or lines
    \item A full FUV census of unobscured stars above a mass limit ($>30$--50~$M_\odot$) across a sample of extremely metal-poor galaxies
    \item FUV-blue optical spectra of entire population of unobscured stars --- representing the first library of empirical spectra in this metallicity and mass range
    \item As described below and in Table~\ref{tab:objectives}: the FUV (particularly 1000--1800~\AA{}) is critical for measuring terminal wind velocities and mass loss rates, metal abundances, as well as luminosities; and the blue optical (particularly 4000--5000~\AA{}) is in turn crucial for measuring surface gravities (and thus spectroscopic masses) and effective temperatures
\end{itemize}

\paragraph{Nebular gas parameters}
(NB: this science effectively comes `for-free' with IFU/IFS observations that accomplish the key stellar spectroscopy described above and simultaneously collect spatially-resolved maps of nebular line emission):
\begin{itemize}
    \item Maps of ionizing photon production at e.g. $>13.6$, $>25$, and $>54$~eV via mapping of nebular line emission probing different ionization potentials across nebulae
    \item Measurement of ionizing photon escape, with implications for feedback models and cosmic reionization (from comparison of total observed nebular emission with atmosphere models constrained by the above stellar spectroscopy; and/or from direct Lyman-continuum constraints at $\lambda<912$~\AA{})
    \item Ionization and electron temperature and density structure across star-forming regions, in spatial relation to ionizing sources and stellar population constraints
    \item Chemical abundances / constraints on recent enrichment across nebulae, linked to spatially resolved recent star formation history
    \item Mechanical energy and other feedback injected into nebulae and detectable in outflow (outflow rates, velocities, chemistry): via both broad gas components detectable in emission and narrow components in absorption in FUV against the backlights of the massive stars themselves \citep[as demonstrated with HST in the Clouds; e.g.][]{hamanowiczMETALZMeasuringDust2024}
    \item As outlined below and in Table~\ref{tab:objectives}: key diagnostics for the nebular gas (both nebular line and nebular continuum emission) span essentially the same range as the key stellar diagnostics: 900--2000~\AA{} in the FUV and 3400--5000~\AA{} in the blue-optical.
\end{itemize}

\section{Galaxy Sample}

\begin{table}
\centering
\tiny
\caption{Nearest Extremely Metal-Poor ($\leq 10\%$~$Z_\odot$) Star-Forming Galaxies \label{tab:galaxies}}
    \begin{tabular}{m{0.8cm}C{0.5cm}C{0.5cm}>{\raggedleft\arraybackslash}p{4.8cm}}
    \hline
        Galaxy & $Z/Z_\odot$ & Distance & Notes and references \\
                                          & (from O/H) & (Mpc) & \\
    \hline\hline
    \grey{Leo~A} & \grey{$1/20$} & \grey{0.8} & \grey{no stars $>30~M_\odot$ \citep{gullPanchromaticStudyMassive2022,urbanejaMetallicityDistanceLeo2023}} \\
    \hline
    \grey{SagDIG} & \grey{$1/20$} & \grey{1.1} & \grey{no stars $>30~M_\odot$ \citep{garciaMassiveStarsSagittarius2018}} \\
    \hline
    \grey{ Sextans~A } & \grey{ $1/10$ } & \grey{ 1.3 } & \grey{ no stars $\gg 40~M_\odot$ \grey{\citep[][Lorenzo+\ in-prep]{lorenzoNewReferenceCatalogue2022,telfordObservationsExtremelyMetalPoor2024}}} \\ 
    \hline
    \grey{Leo P} & \grey{$1/30$} &  \grey{1.6} & \grey{no stars $>30~M_\odot$ \citep{evansFirstStellarSpectroscopy2019,telfordFarultravioletSpectraMainsequence2021}} \\
    \hline
    \multicolumn{4}{c}{\emph{Resolved star distance limit for HST and other current facilities}} \\
    \hline
    UGC~8091 & $1/10$ & 2.2 & \citet{bergDirectOxygenAbundances2012} \\
    \hline
    UGC~6817 & $1/15$ & 2.7 & \citet{bergDirectOxygenAbundances2012} \\
    \hline
    NGC~4163 & $1/15$ & 3.0 & \citet{bergDirectOxygenAbundances2012} \\
    \hline
    J1258+1413 & $1/25$ & 3.0 & \citet{izotovLowredshiftLowestmetallicityStarforming2019} \\
    \hline
    NGC~3741 & $1/10$ & 3.2 & \citet{bergDirectOxygenAbundances2012} \\
    \hline
    UGC~4483 & $1/15$ & 3.6 & \citet{bergDirectOxygenAbundances2012} \\
    \hline
    UGCA~292 & $1/25$ & 3.9 & \citet{bergDirectOxygenAbundances2012} \\
    \hline
    UGC~7605 & $1/10$ & 4.4 & \citet{bergDirectOxygenAbundances2012} \\
    \hline
    CGCG~269-049 & $1/15$ & 4.6 & \citet{bergDirectOxygenAbundances2012} \\
    \hline
    Peekaboo / HIPASS J1131-31 & $1/50$ & 6.8 & \citet{karachentsevPeekabooExtremelyMetal2022,kniazevPeekabooGalaxyNew2025} \\
    \hline
    UGC~4278 & $1/10$ & 7.6 & \citet{bergDirectOxygenAbundances2012} \\
    \hline
    J0940+2935 & $1/10$ & 8 & \citet[]{izotovMMTObservationsNew2007,senchynaExtremelyMetalpoorGalaxies2019} \\
    \hline
    J0959+4626 & $1/30$ & 8.7 & \citet{izotovLowredshiftLowestmetallicityStarforming2019} \\
    \hline
    UGC~3600 & $1/10$ & 9.3 & \citet{pustilnikStudyGalaxiesLynxCancer2016} \\
    \hline
    MCG9-13-56 & $1/10$ & 10.0 & \citet{pustilnikStudyGalaxiesLynxCancer2016} \\
    \hline
    KKH~46 & $1/10$ & 10.0 & \citet{pustilnikStudyGalaxiesLynxCancer2016} \\
    \hline
    UGC~3501 & $1/15$ & 10.1 & \citet{pustilnikStudyGalaxiesLynxCancer2016} \\
    \hline
    J0859+3923 & $1/20$ & 10.2 & \citet{izotovLowredshiftLowestmetallicityStarforming2019} \\
    \hline
    KISSB~23 & $1/10$ & 10.2 & \citet{pustilnikStudyGalaxiesLynxCancer2016} \\
    \hline
    UGC~5272B & $1/10$ & 10.3 & \citet{pustilnikStudyGalaxiesLynxCancer2016} \\
    \hline
    UGC~695 & $1/10$ & 10.5 & \citet{kniazevStudyGalaxiesEridanus2018} \\
    \hline
    J0926+3343 & $1/35$ & 10.6 & \citet{pustilnikStudyGalaxiesLynxCancer2016} \\
    \hline
    J0911+3135 & $1/20$ & 10.8 & \citet{izotovLowredshiftLowestmetallicityStarforming2019} \\
    \hline
    UGC~521 & $1/10$ & 10.9 & \citet[]{bergDirectOxygenAbundances2012} \\
    \hline
    HS~1442+4250  & $1/10$ & 11 & \citet[]{gusevaSpectroscopicPhotometricStudies2003,senchynaExtremelyMetalpoorGalaxies2019} \\
    \hline
    UGC~9497 & $1/20$ & 11.0 & \citet{gusevaSearchingMetaldeficientEmissionline2017} \\
    \hline
    J0812+4836 & $1/26$ & 11.1 & \citet{pustilnikStudyGalaxiesLynxCancer2016} \\
    \hline
    Leoncino & $1/40$ & 12.1 & \citet{mcquinnLeoncinoDwarfGalaxy2020} \\
    \hline
    \textbf{DDO~68} & \textbf{1/30} & \textbf{13} & $\mathbf{\mathrm{\textbf{SFR}}\simeq 0.2~M_\odot/\mathrm{yr}}$ \textbf{\citep[e.g.][]{pustilnikStudyDDO682005,makarovUnusualVoidGalaxy2017,sabbiResolvedStellarPopulations2018}} \\
    \hline
    HS~0822+3542 & $1/22$ & 13.5 &  \citet{kniazevHS082235422000,pustilnikStudyGalaxiesLynxCancer2016} \\
    \hline
    J0955+6442 & $1/30$ & 13.9 & \citet{izotovLowredshiftLowestmetallicityStarforming2019} \\
    \hline
    J0927+0314 & $1/15$ & 15 & \citet{senchynaPhotometricIdentificationMMT2019} \\
    \hline
    J1119+0935 & $1/30$ & 16 & \citet{izotovLowredshiftLowestmetallicityStarforming2019} \\
    \hline
    J0113+0052 & $1/30$ & 15.8 & \citet{izotovLowredshiftLowestmetallicityStarforming2019} \\
    \hline
    J1005+3722 &  $1/30$ & 15.9 & \citet{senchynaPhotometricIdentificationMMT2019,brenemanLeonessaExtremelyMetalpoor2025} \\
    \hline
    J1226+0952 & $1/30$ & 16.2 & \citet{izotovLowredshiftLowestmetallicityStarforming2019} \\
    \hline
    MCG+00-04-049 & $1/15$ & 17.2 & \citet{kniazevStudyGalaxiesEridanus2018} \\
    \hline
    J1109+2007 & $1/30$ & 17.7 & \citet{izotovLowredshiftLowestmetallicityStarforming2019} \\
    \hline
    \textbf{\izwi{}} & \textbf{1/32} & \textbf{18.2} & $\mathbf{\mathrm{\textbf{SFR}}\simeq 1 M_\odot/\mathrm{\textbf{yr}}}$ \textbf{\citep[e.g.][]{aloisiZw18Revisited2007,annibaliStarFormationHistory2013}}: \textbf{\textit{hundreds--thousands of O stars depending on detailed SFH}} \\
     \hline
    HS~1013+3809 & $1/10$ & 19.3 & \citet{pustilnikStudyGalaxiesLynxCancer2016} \\
    \hline
    Little Cub & $1/36$ & 20.6 & \citet{hsyuLittleCubDiscovery2017} \\
    \hline
\end{tabular}
\end{table}
Hundreds of extremely metal-poor star-forming galaxies host to large populations of stars above 30~$M_\odot$ stars have been identified in the local Universe \citep[e.g.][]{kunthMostMetalpoorGalaxies2000,gusevaSearchingMetaldeficientEmissionline2017}; but critically, none reside nearby enough for current facilities to spatially resolve individual massive stars that could be analyzed spectroscopically.
Pathfinding programs with the largest telescopes on Earth and HST, targeting the small number of $\lesssim 1$~Mpc dwarf irregulars with any extremely metal-poor massive stars, hint at notable differences from their higher-metallicity counterparts: including extremely weak winds and signs of fast rotation and potential binary interaction \citep[e.g.][Senchyna+\ in-prep]{garciaLowmetallicitySubSMCMassive2017,telfordFarultravioletSpectraMainsequence2021,gullPanchromaticStudyMassive2022,lorenzoNewReferenceCatalogue2022,telfordObservationsExtremelyMetalPoor2024}.
But none of these programs have identified any extremely metal-poor stars significantly above $40$~$M_\odot$, from which the bulk of observed UV continuum and ionizing light originate in very young starbursts at high-redshift. 
The key problem is that the recent star formation histories of these closest extremely metal-poor galaxies are simply too low to efficiently sample the most massive and shortest-lived OB stars.

We compile in Table~\ref{tab:galaxies} a list of the nearest galaxies known with 1) extremely metal-poor gas ($\lesssim 10\%$ solar in oxygen) and 2) evidence for an unobscured massive star population (for reference, galaxies for which we can access massive stars now with HST but which don't host significant numbers of stars in the targeted mass and metallicity range are noted in \grey{grey}).
These galaxies are collated from several sources with differing methodologies \citep[][]{bergDirectOxygenAbundances2012,pustilnikStudyGalaxiesLynxCancer2016,gusevaSearchingMetaldeficientEmissionline2017,izotovLowredshiftLowestmetallicityStarforming2019,brenemanLeonessaExtremelyMetalpoor2025}; future work obtaining uniform star formation history and distance measurements for this sample will further narrow down the candidate list of those most likely to host the most substantial massive star populations for follow-up.
Optical imaging and nebular spectroscopy suggests that the majority are very low-mass and have low SFRs \citep[$\lesssim 10^{-2}$~$M_\odot/\mathrm{yr}$; e.g.][]{senchynaExtremelyMetalpoorGalaxies2019} over the past 10~Myr; and thus are likely host to few stars $>30$--$50$~$M_\odot$.
We note that this potential target list is not final; in particular, UVEX is poised to deliver a much more complete inventory of star-forming dwarfs with deep UV all-sky photometry that may uncover other promising nearby objects \citep{kulkarniScienceUltravioletExplorer2021}. 
However, given the depth of past objective prism surveys, which are effective at detecting metal poor starbursting dwarf galaxies \citep{jclee00, jclee04}, this is unlikely to substantially change the prime target list for study with HWO.

Only a handful of galaxies are likely to host sizable populations of massive stars significantly below 10\% solar metallicity; and all reside beyond $\sim$5--10 Mpc.
As highlighted below, this places them out of range of current facilities.
Arguably the most critical of these systems is \izwi{}: which hosts the highest SFR at $\sim 1~M_\odot/\mathrm{yr}$ during the past 10~Myr \citep{annibaliStarFormationHistory2013} and among the lowest metallicities ($\sim 1/32~Z_\odot$) of extremely metal-poor galaxies within 20~Mpc.
A facility that places \izwi{} at 18.2~Mpc in range of resolved star work would be capable of conducting similar observations extending to fainter stars and to higher-SNR over a sample of tens of other galaxies (see Table~\ref{tab:galaxies}) more nearby but less extreme in recent star formation activity and metallicity.
While observations of \izwi{} and other systems at $\gtrsim 10$~Mpc would provide our first look at the critical top of the IMF ($>50~M_\odot$) at the lowest metallicities, observations with the same facility will also revolutionize our view of detailed massive star evolution at $\sim10\%$~$Z_\odot$ by providing spectroscopy of similar quality to that currently achievable in the Magellanic Clouds to the lower-SFR dwarf Irregular galaxies (in particular, Sextans A and Leo A) at and just beyond the edge of the Local Group (see Table~\ref{tab:galaxies} and Appendix~\ref{app:relatedscience}).

\izwi{} is a particularly important target for many reasons.
Integrated spectra with COS confirm that it powers nebular emission in highly-ionized \ion{C}{4} and \ion{He}{2} \citep[e.g.][]{mingozziCLASSYIVExploring2022}, lines which are commonly encountered and still puzzled-over at high-redshift.
In addition, several UV and optical studies have detected signatures of what appears to be broad stellar wind emission typically attributed to Wolf Rayet stars or similar atmospheres in parts of the galaxy \citep[e.g.][]{izotovZw18New1997,legrandDetectionWRStars1997,brownIsolatingClustersWolfRayet2002}.
This is extremely intriguing and puzzling, as such stars are not expected to be produced at all in the single star evolutionary paradigm at such low metallicities, let alone power such strong winds \citep[e.g.][]{crowtherReducedWolfRayetLine2006} unless they are very massive \citep[e.g.][]{vinkVeryMassiveStars2018}.
The detection of these lines would suggest that \izwi{} is host to very luminous stars close to the Eddington limit, potentially of exceptional initial mass or perhaps products of binary mass transfer or chemically homogeneous evolution \citep{szecsiLowmetallicityMassiveSingle2025}.
Follow-up deep optical spectroscopy has not resolved this conundrum \citep[e.g.][]{kehrigExtendedHeII2015,rickardsvaughtKeckCosmicWeb2021}, highlighting the importance of high spatial resolution UV work not possible with current instrumentation.
Only HWO and resolved observations in the UV/blue optical are capable of unraveling this mystery and taking advantage of the opportunity that the potential presence of such stars represents.

\begin{figure*}
    \centering
    \includegraphics[width=1.0\textwidth]{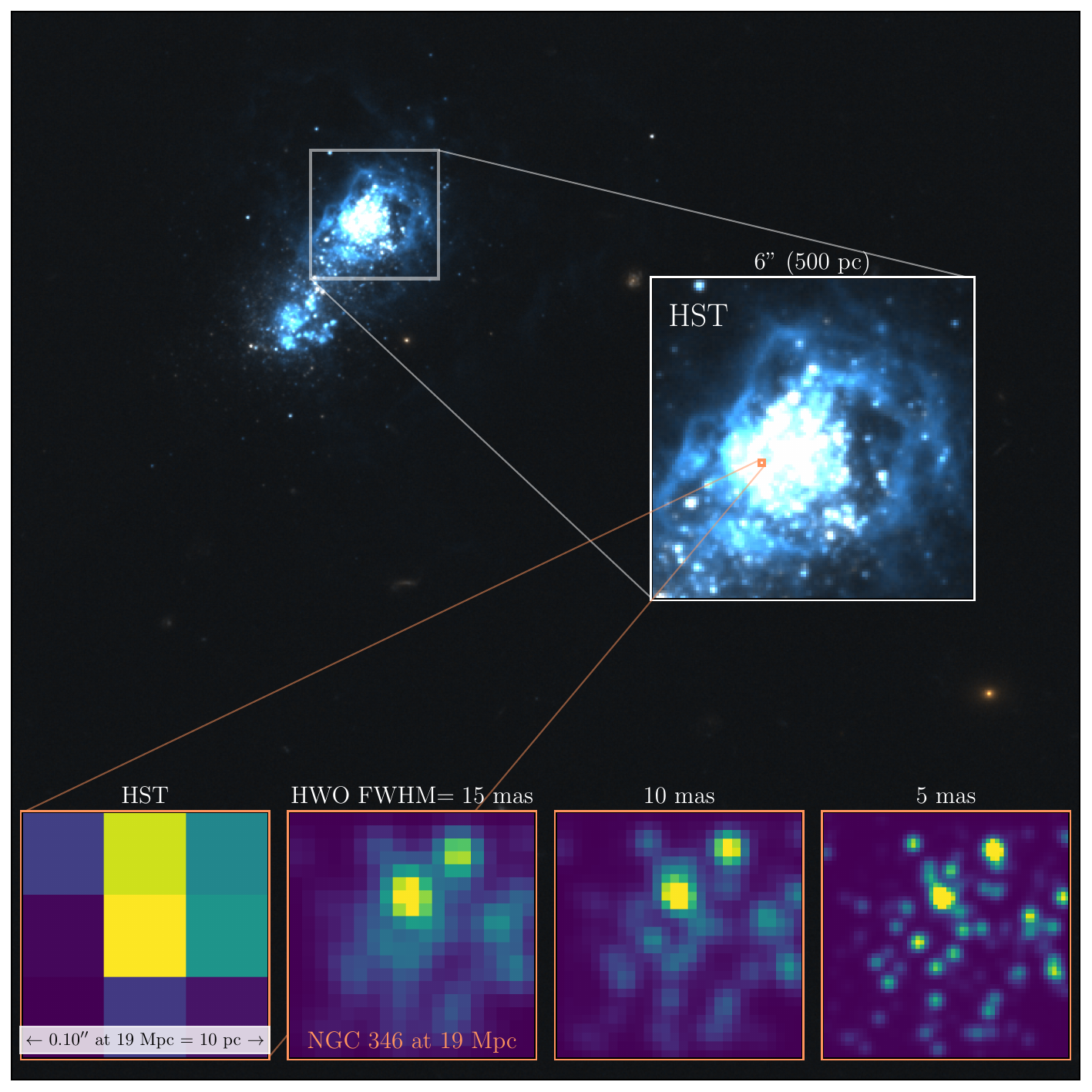}
    \caption{
        A particularly important target is the well-known extremely metal-poor galaxy I~Zw~18 -- the nearest bastion of a substantial population of massive ($>30$~$M_\odot$) stars significantly below 10\% solar metallicity (HST/ACS~F606W$+$F814W imaging from GO:10586, PI:Aloisi).
        The core of the NW component of the galaxy is a cluster complex subtending $4''$ on a side ($\sim 300$~pc at the 18~Mpc distance of I~Zw~18).
        At bottom, we display a simulation of a zoom-in on a 10~pc region of this cluster complex with different UV PSFs, ranging from that achieved by HST/WFC3 ($\sim 80$~mas) down to 5~mas.
        This 10~pc region subtends a very small area of sky: $0.10''$ at 18 Mpc (the approximate size of just 1 JWST microshutter).
        We display for reference the approximate pixel size of HST/WFC3/UVIS relative to this scale (left).
        To produce a simulated view from HWO, we take a cutout of an HST F275W (NUV) image of the core of the most massive star-forming region in the SMC, NGC 346 (GO:17118, PI:Murray).
        We plot versions of this NGC 346 image convolved to different prospective HWO PSF sizes (and resampled at Nyquist) as labeled.
        With a UV(-blue optical) IFU (or alternatively, a microshutter array of similar spatial scale and allowing for similarly dense multiplexing) sampling spatial scales $\lesssim 15$~milliarcseconds (mas) over a field of view of $\geq 0.1''$ (ideally, FOV$\sim 3$--$10''$), HWO is uniquely capable of resolving and taking the critical UV spectra of a large population of individual massive $\gtrsim 30$~$M_\odot$ stars within the NW star-forming complex of I~Zw~18 (with the actual expected yield of resolved stars scaling inversely with the achieved PSF FWHM).
    \label{fig:izw18demo}
    }
\end{figure*}

\begin{figure}[ht!]
    \centering
    \includegraphics[width=0.5\textwidth]{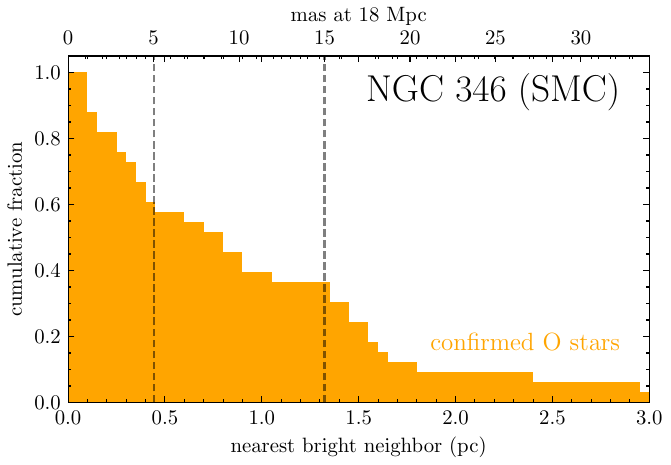}
    \caption{
        As one empirical estimate of crowding in \izwi{}, we examine the cumulative distribution of distance to the nearest bright neighbor ($m_V<18$, $M_V<-1$) for O stars \citep[spectroscopically-confirmed:][]{evansVLTFLAMESSurveyMassive2006,duftonCensusMassiveStars2019} in the cluster NGC~346 in the SMC, cross-matching to the HST photometric catalog of \citet[][and since this neighbor comparison is done at $\sim 5000$~\AA{}, this represents a conservative approximation of crowding at FUV wavelengths]{sabbiPresentStarFormation2007}.
        A PSF and spectroscopic aperture $<20$~mas would begin to resolve some O stars in a similar cluster at 18~Mpc; with the fraction resolved scaling strongly with increasingly finer spatial sampling up to approximately half or higher at $<15$~mas.
        % This provides one estimate of crowding in \izwi{}; some regions are likely to be less densely concentrated than in this cluster, allowing for a higher effective yield
    \label{fig:ngc346crowding}
    }
\end{figure}

Key to the science described here is the ability to spatially-resolve and take spectra of individual massive stars in these extremely metal-poor environments.
This is well beyond the reach of current facilities (Fig.~\ref{fig:izw18demo}).
Resolving a substantial fraction of IZw18's massive star content is plausible with HWO. HST imaging suggests that its massive stars are not all as densely-packed as in the prototypical dense young cluster R136 and instead may resemble local open clusters in compactness \citep{hunterMassiveStarsZW1995,annibaliStarFormationHistory2013}.
While extremely dense cores of clusters in IZw18 will not be resolvable with a $<10$~m aperture telescope \citep[e.g.][]{garciaMassiveStarsExtremely2021}, a significant fraction of its massive stars could lie within reach of such a telescope.
To estimate the ability of HWO to dissect this population, we leverage data from the SMC open cluster NGC~346 (Fig.~\ref{fig:ngc346crowding}).
A facility with PSF and spectroscopic sampling on the order of $\lesssim 15$–$20$~mas would open the door to resolving O stars in a comparable cluster in IZw18 at 18~Mpc and obtaining high-quality spectra of the most densely packed subset.

The key advantage of HWO is that it would allow a substantial fraction of the young stellar component of \izwi{} and similar galaxies to be resolved.
At the same time, observations at lower angular resolution would remain highly valuable, providing critical insights that address a complementary set of questions but which would rely more heavily on analysis with parameterized population synthesis models of blended populations (Table~\ref{tab:resolutionimpacts}).
% \begin{figure*}[ht!]

\section{Description of Observations}
\label{sec:observations}
\begin{table*}[ht!]
    \centering
    \footnotesize
    \caption[Performance Goals]{Constraints on Metal-Poor Massive Star Physics as a Function of Science Return}
    \label{tab:resolutionimpacts}
    \begin{tabular}{m{2.0cm}|C{3.2cm}C{3.2cm}C{3.2cm}C{3.2cm}}
        \hline
        & {Current State of the Art} & {Incremental Progress} & {Substantial Progress} & {Major Progress / \emph{The Dream}} \\
        \hline
        \hline
        {Type of analysis} & Comparison of integrated stellar+nebular galaxy light to population synthesis models  & Comparison of integrated clusters+nebular emission to population synthesis models & Semi-resolved clusters: some resolved stars/clumps, spatially-resolved high-ionization nebular structure & Resolved stars and compact associations within clusters; \emph{The Dream: majority of O-stars resolved at 18~Mpc} \\
        \hline
        Degeneracies & star formation histories, IMF, unresolved gas physics  & cluster ages, IMF & contents of blended clumps and clusters & contents of the most crowded cluster regions \\
        \hline
    \end{tabular}
\end{table*}

\begin{table*}[ht!]
    \centering
    \footnotesize
     \caption[Observation Requirements]{Capabilities Needed to Enable Different Levels of Scientific Return}
    \label{tab:thedream}
    \begin{tabular}{m{3.0cm}|C{2.5cm}C{2.2cm}C{2.2cm}C{2.2cm}C{2.5cm}}
        \hline
        {} & {Current State of the Art} & {Incremental Progress} & {Substantial Progress} & {Major Progress} & \emph{The Dream} \\
        \hline
        \hline
        Observation type for extremely low-$Z$ massive stars & Imaging, slit/fiber spectroscopy [\emph{disparate instruments, low sensitivity}]  & UV MOS & {\centering UV IFU \\ {[fallback: MOS]}} & UV--optical IFU [fallback: MOS] & UV--optical IFU \\
        \hline
        Wavelength range & panchromatic & 900--2000~\AA{} & 900--2000~\AA{} & 900--2000, 3000--5000~\AA{}  & 900--5000~\AA{} \\
        \hline
        Spectral resolution & R$\sim$16000 (highest achieved in UV with HST/COS) & $\mathcal{R}=3000$ & $\mathcal{R}=3000$ & $\mathcal{R} \gtrsim5000$ & $\mathcal{R} \gtrsim5000$ \\
        \hline
        Angular resolution/spaxel size for spectroscopy & $\gtrsim 0.05''$ (smallest slit size available in the UV with HST/STIS [with significant losses]) & $0.05''$ & $0.03''$ & $\lesssim 0.02''$ & {\centering $\sim 0.005''$ \\ (8-m diffraction-limit at 1500~\AA{})} \\
        \hline
        Spatial resolution for massive stars at 18~Mpc (I~Zw~18) & $\gtrsim 5$~pc & $ 5$~pc & $3$~pc & $\lesssim 2$~pc & 0.5~pc\\
        \hline
        Fraction of NGC~346 O-stars resolved at 18~Mpc & $\sim 0\%$ & $\sim 0\%$ & 5\% & $\gtrsim 10\%$ & $60\%$ \\
        \hline
    \end{tabular}
\end{table*}

\subsection{Stellar spectroscopy}

\emph{Target dataset for stellar spectroscopy:} FUV--optical spectroscopy for the first sample ($N \gg 10$) of spatially resolved extremely metal-poor massive stars $\gtrsim 30~M_\odot$, at $\mathrm{NUV}<25$ (absolute mag $\lesssim -6$) stars at $\mathcal{R}>4000$ and $\mathrm{SNR}>10$ over 1--10 $Z<10\%~Z_\odot$ dwarf galaxies.

This demands either a UV(--blue optical) IFU operating with spatial scale approaching the diffraction limit over a FOV $>0.1''$ (ideally, $\gtrsim 3''$; see Fig.~\ref{fig:izw18demo}); or a MOS with similarly stringent spatial sampling (with significant efficiency lost due to the inefficient multiplexing).
This is a critical requirement; massive stars in these systems will not be individually resolvable with any confidence if the characteristic spatial scales probed by the IFU or slits are $\gg0.020''$ (Fig.~\ref{fig:ngc346crowding}), significantly changing the character of the science possible.

In the (ideal) IFU case, the necessary spectroscopy for both the stars and gas (see Table~\ref{tab:thedream} below) are collected at once.

In the finely-sampled MOS case, the nebular science would be limited; and a multistep observing strategy would be required to collect the stellar spectroscopy:
\begin{itemize}
    \item Deep FUV--NUV--optical imaging of target galaxies for spectroscopic target pre-selection (as discussed above, the stars in the densest regions of \izwi{} and our other targets are not resolved in imaging by any existing facility)
    \begin{itemize}
        \item FUV--optical, to $\sim25$--30 mag (key: must detect all unobscured massive stars down to $\sim 20~M_\odot$). UV critical for FUV spectroscopy planning, and photometric estimates of $T_\mathrm{eff}$/$L_\mathrm{bol}$/extinction
    \end{itemize}
    \item Moderate-resolution follow-up spectroscopy spanning FUV--blue optical with the MOS (likely requiring an order of magnitude larger time expenditure due to multiple configurations necessary to obtain spectra of a large sample of closely-spaced targets.)
    \item Alternatively, slit-stepping with a sufficiently narrow slit and well-behaved PSF could emulate an IFU observation; but at significantly lower observational efficiency.
\end{itemize}

In either case, the spectroscopic requirements are similar (see also Table~\ref{tab:objectives}):
\begin{itemize}
    \item Critical need for UV: 
    \begin{itemize}
        \item To detect and study the hottest stars ($\gg 40$~kK) including binary products / stripped stars / fast rotators, some of which (e.g.\ stripped stars) are bolometrically faint and easily outshined by companions at longer wavelengths \citep[e.g.][]{gotbergIonizingSpectraStars2017} 
        \item To access wind signatures: \ion{O}{6}~$\lambda \lambda 1032,1038$ \citep[important X-ray sensitive line, with implications for mass loss rates in highly-ionized winds; e.g.][]{bianchiEffectiveTemperaturesMidO2002, zsargoImportanceInterclumpMedium2008}, \ion{S}{4}~$\lambda\lambda 1062,1072$, \ion{P}{5} $\lambda\lambda 1118, 1128$, \ion{N}{5} $\lambda\lambda 1238,1242$, \ion{O}{5} $\lambda 1371$, \ion{C}{4} $\lambda\lambda 1548,1550$, \ion{N}{4} $\lambda 1718$) sensitive to terminal velocity ($v_\infty$) and mass loss rate ($\dot{M}$)
        \item To study nebular gas emission lines tracing highly-ionized, hot gas; including \ion{N}{4}] $\lambda\lambda 1483,1486$,  \ion{C}{4} $\lambda \lambda 1548,1550$, \ion{He}{2} $\lambda 1640$, \ion{O}{3}] $\lambda \lambda 1661,1666$, [\ion{C}{3}], \ion{C}{3}] $\lambda \lambda 1907,1909$. Crucial diagnostics of high energy feedback processes and routinely detected in integrated galaxy spectra at high redshift.
        \item To correctly model atmospheres for inference of fundamental parameters and wind/ionizing feedback for stars with (uncertain) weak winds expected at low-metallicity \citep[e.g.][]{kudritzkiLinedrivenWindsIonizing2002,vinkTheoryDiagnosticsHot2022}
        \item To access Fe/H-sensitive (especially those at 1400--1540~\AA{}) and CNO-sensitive photospheric features
        \item To detect FUV absorption lines from interstellar gas at a range of ionization states in outflow and inflow seen in-projection
    \end{itemize}
    \item And of blue-optical:
        \begin{itemize}
            \item Critical diagnostic lines of temperature and gravity reside exclusively in the blue optical, $\sim 3000$--5000~\AA{} \citep[e.g.][]{walbornContemporaryOpticalSpectral1990,martinsUVOpticalNearIR2011,simon-diazModernGuideQuantitative2020,evansNearUVReconnaissanceMetalpoor2023}.
        \end{itemize}
\end{itemize}

\paragraph{Spectral resolution:}
\begin{itemize}
    \item Basic absolute minimum requirement on $\mathcal{R}$ is $>2000$ to separate interstellar from stellar absorption in photospheric and wind profiles and measure terminal velocities (slightly more stringent minimum than quoted by \citealt{crowtherReliabilityCIVl1549Abundance2006} to compensate for lower metallicity); $>4000$ is required for robust extraction of velocity profiles and recovery of faint wind lines at the slow wind speeds anticipated at such low metallicity.
    \item While much higher resolution in the UV is currently possible with the medium resolution gratings on HST/COS (Table~\ref{tab:thedream}), this is not required for typical wind or photospheric line extraction; and a more moderate resolution of $R\sim 4000$--8000 provides a reasonable balance with the required SNR and wavelength range.
    \item In the optical, extracting surface gravity and rotational velocity constraints requires good signal-to-noise at $\mathcal{R}\gtrsim 4000$--8000
\end{itemize}

\paragraph{Sensitivity / exposure time:}
\begin{itemize}
    \item Accounting for modest extinction ($A_V=0.1$) and the 18.2~Mpc distance from \citet{aloisiZw18Revisited2007}, MIST tracks for massive stars at the expected few-percent solar predict stars $>30~M_\odot$ ($>50~M_\odot$) have FUV (GALEX) mags $<26$~AB ($<25$~AB). Approximate exposure times according to the current UV spectroscopic ETC suggest $\mathrm{SNR}=10$ at $\mathcal{R}=5000$ can be reached in $\lesssim 100$ hours for stars $<26$ for an 8m configuration. The most massive stars $\gtrsim 100~M_\odot$ at $<24$~AB require significantly shorter exposures. Future work will explore how this scales with detailed telescope configuration and more robustly model spectral feature recovery from the UV to the blue-optical.
    \item \emph{Efficient multiplexing is critical}: a single IFU pointing with small FOV $\sim3''$ and high enough spatial resolution would resolve and acquire spectra for a large sample $\mathcal{O}(100)$ of massive stars (and their surrounding ionized gas) in e.g.\ the NW cluster of \izwi{}, accomplishing our key science in a single deep dwell.
\end{itemize}

\paragraph{Spatial resolution:}

This is the key that enables the key science return of resolved stellar science. 
As a first step towards understanding the impact of spatial resolution, we refer the reader back to Fig.~\ref{fig:ngc346crowding}, which can be interpreted as a first-order empirical estimate of a `yield' of resolved stars from a cluster population similar to that in NGC~346 as a function of achieved PSF and sampling scale.
The number of stars resolved and thus our ability to populate an HR diagram and constrain model predictions for stars over a broad mass and evolutionary range scales strongly with increasingly finer spatial sampling.

\subsection{Nebular spectroscopy}

\emph{Target dataset for nebular spectroscopy:} FUV--optical IFU observations of scale $\sim 5''\times 5''$ at $\mathcal{R}>4000$ over $Z<10\%~Z_\odot$ dwarf galaxies

\begin{itemize}
    \item  Similar but in principle slightly less stringent spatial resolution considerations than the coupled stellar science above --- comparable spatial resolution is critical to tying these gas conditions to the stars/clusters responsible
    \begin{itemize}
        \item Broad spectral range: $\sim 1200$--5000 (Ly$\alpha$-[\ion{O}{3}]) captures the key diagnostics of both high and low ionization nebular gas
        \item High resolution: $\mathcal{R}\gtrsim 4000$ critical to constraining broad line components at virial scale ($\gtrsim 50$--100~km/s) and to disentangling different contributions to UV lines (ISM absorption / nebular emission / stellar absorption/emission)
        \item Critical need for UV:
        \begin{itemize}
            \item Access to high-ionization lines including \ion{C}{4} and \ion{He}{2} sensitive to hottest ionizing sources and densest gas; and which are directly observed in integrated light in highest-redshift galaxies
            \item Access to unique CNO enrichment signatures: \ion{C}{3}], \ion{C}{4} / \ion{O}{3}] and \ion{N}{3}], \ion{N}{4}] (+[\ion{N}{2}]) / \ion{O}{3}] constrain C/O and N/O in densest / hottest gas around massive stars
            \item Access to ionized outflow absorption signatures complementary to emission signatures in optical
        \end{itemize}
    \end{itemize}
\end{itemize}

Note that this science cannot be accomplished by the ELTs.
While the ELTs operating with AO will resolve comparable spatial scales in the near-infrared over small fields of view ($\lesssim 1''$), the best-studied diagnostic lines necessary for physical characterization of hot stars and their feedback reside in the UV-blue optical \citep[e.g.][]{evansELTSpectroscopyExtragalactic2018}; and crucially, hot massive stars are too faint at IR wavelengths to observe at the requisite fidelity at Mpc-distances.
The lack of diffraction-limited spectrophotometry at short wavelengths means that science with the ELTs in this space will only provide at best a rough preview of the hot stars that a facility like HWO is necessary to constrain in detail \citep[see also e.g.][]{garciaMassiveStarsExtremely2021}.

\appendix
\section{Related Science with HWO}
\label{app:relatedscience}
While it is beyond the scope of this particular SCDD (which is aimed primarily at high mass ($> 30~M_\odot$) and extremely metal-poor ($<10\%$~$Z_\odot$) massive stars and as a result is focused on a high spatial resolution IFS), HWO will also enable a dramatic expansion of science possibilities in other metal-poor stellar populations.

\paragraph{Detailed stellar astrophysics at ~10\% solar with a wide-field MOS:}
A wide-field ($\sim$arcmins$^2$) sensitive UV--optical MOS on HWO would have transformational impact on our understanding of metal-poor stellar evolution at $\lesssim 40~M_\odot$ by dissecting the population of $\sim 1/10~Z_\odot$ dwarfs at $\sim 1$~Mpc (in particular, Sextans~A and Leo~A) which span larger areas on the sky (4--12 arcmin$^2$) at significantly lower source densities.
UV spectroscopy with such an instrument would yield the first thorough characterization of stellar winds, photospheric chemistry, and binary demographics in these systems.
At the moment, the samples of stars accessible with HST are restricted to just of-order a dozen of the brightest stars, each of which requires many orbits to obtain even a single low-resolution snapshot of the UV.
And while clues from this spectroscopy hint at significant deviations from expectations in wind strengths, rotation rates, and binary products \citep[e.g.][]{gullPanchromaticStudyMassive2022,telfordFarultravioletSpectraMainsequence2021,telfordObservationsExtremelyMetalPoor2024}, population-level conclusions are severely limited by sample size and data quality.

Extension of such an instrument to optical wavelengths would also enable unparalleled observations where ground-based, 10m-class telescopes are meeting their sensitivity limit.
The large collecting power of future ELTs will only provide a partial solution because of the planned instrumentation and the limitations of adaptive optics. For instance, the ELT wide-field spectrograph MOSAIC will be fed by $\sim$200~mas fibers, hence multiple sources may enter the fibers in the most crowded regions.
Moreover, the wide separation between fibers in the focal plane will not enable effectively covering complete galaxies.
The high spatial resolution IFU HARMONI can overcome crowding issues but its very small FoV is again impractical to cover full populations of $\sim 1$~Mpc galaxies; and it will only reach the brightest objects in the $J$-band with sufficient SNR for detailed characterization.
A wide-field MOS on HWO would naturally overcome spatial resolution issues and provide the broad spatial coverage needed to characterize the winds and detailed evolution of $1/10~Z_\odot$ massive stars.

{\bf Acknowledgements.} 

Based on observations made with the NASA/ESA Hubble Space Telescope, and obtained from the Hubble Legacy Archive, which is a collaboration between the Space Telescope Science Institute (STScI/NASA), the Space Telescope European Coordinating Facility (ST-ECF/ESA) and the Canadian Astronomy Data Centre (CADC/NRC/CSA).

PS acknowledges support associated with program \#17129 provided by NASA through a grant from the Space Telescope Science Institute, which is operated by the Association of Universities for Research in Astronomy, Inc., under NASA contract NAS 5–26555.
MG gratefully acknowledges support by grants PID2022-137779OB-C41 and PID2022-140483NB-C22, 
funded by % the Spanish Ministry of Science, Innovation and Universities/State Agency of Research 
MICIU/AEI/10.13039/501100011033 and by “ERDF A way of making Europe”,
and grant MAD4SPACE, TEC-2024/TEC-182 from Comunidad de Madrid (Spain).

\bibliography{zoterolib.bib, author}

\end{document}